\documentclass[12pt]{article}
\usepackage{amsmath,amsfonts,xcolor,amsthm,natbib,prodint,graphicx}
\usepackage[margin=1in]{geometry}
\usepackage{hyperref}
\usepackage[capitalize,nameinlink]{cleveref}

\newcommand{\E}{\mathbb{E}}
\renewcommand{\P}{\mathbb{P}}

\newcommand{\R}{\mathbb{R}}

\newcommand{\ind}{\perp\!\!\!\!\perp}
\renewcommand{\d}{\mathrm{d}}
\newcommand{\M}{\mathcal{M}}

\newcommand{\bbeta}{\boldsymbol{\beta}}
\newcommand{\bL}{\boldsymbol{L}}
\newcommand{\bl}{\boldsymbol{\ell}}
\newcommand{\bzero}{\boldsymbol{0}}

\newcommand{\balpha}{\boldsymbol{\alpha}}

\newcommand{\bS}{\boldsymbol{S}}

\newcommand{\card}[1]{\lvert #1\rvert}
\newcommand{\logit}{\operatorname{logit}}
\newcommand{\mJ}{\mathcal{J}}

\newtheorem{asp}{Assumption}
\crefname{asp}{assumption}{assumptions}
\Crefname{asp}{Assumption}{Assumptions}
\newtheorem{contexmp}{Example}

\newtheorem{discexmp}{Example}

\crefname{contexmp}{example}{examples}
\Crefname{contexmp}{Example}{Examples}
\crefname{discexmp}{example}{examples}
\Crefname{discexmp}{Example}{Examples}
\newtheorem{prop}{Proposition}
\newtheorem{thm}{Theorem}
\newtheorem{lem}{Lemma}
\newtheorem{cor}{Corollary}
\newtheorem{defn}{Definition}

\newtheorem*{rem}{Remark}

\title{Semiparametric efficient estimation of the Cox regression coefficient when there can be ties}

\usepackage{authblk}
\author{Benjamin R. Baer\footnote{corresponding author: \href{mailto:b.baer@bowdoin.edu}{b.baer@bowdoin.edu}.}}
\affil{Department of Mathematics, Bowdoin College, Maine, USA}

\begin{document}

\maketitle

\begin{abstract}
    The Cox model and the Cox regression coefficient are widely used in applied work, although the corresponding statistical
    theory is principally developed in the continuous case where there can be no tied failure times. 
    In this work, we derive the efficiency bound of the Cox regression coefficient in the Cox model, without imposing
    continuity or discreteness assumptions on the failure distribution. 
    Then, under a technical assumption that the 
    population failure mass points belong to an unknown finite set, we show that the ``exact'' estimator in 
    \citet{cox1972regression} is asymptotically efficient. Next, we propose an asymptotically equivalent 
    estimator which solves the efficient score which has lower computational complexity than the exact Cox score. 
    The development employs recently introduced martingale theory, and throughout examples of the general theory are
    given for both the absolutely continuous case and the discrete case. 
\end{abstract}

\tableofcontents

\section{Introduction}
\label{sec:intro}

Consider time-to-event data consisting of a possibly right-censored event time and baseline covariates. 
The Cox model is the predominant regression model for these data, describing the effect of the
baseline variables through a finite-dimensional regression coefficient while leaving the baseline failure distribution
unrestricted \citep{cox1972regression}. The regression coefficient has maintained a central role in applied work for
more than half a century \citep{kalbfleisch2023fifty}.

The asymptotic behavior of the Cox regression coefficient has received significant attention; however, much of the development
has assumed that the failure distribution is continuous, and often absolutely continuous, thereby excluding mass points and tied
observations. 
Following informal arguments in \citet{cox1975partial}, consistency and asymptotic normality were rigorously established by 
\citet{tsiatis1981large,andersen1982cox}. 
\citet{tsiatis1981large} assumed both failure and censoring were absolutely continuous, while \citet{andersen1982cox} allowed 
the baseline cumulative hazard to be continuous without requiring absolute continuity, but did not allow it to have jumps.
The efficiency of the Cox estimator was highlighted by \citet{cox1972regression} as a major unresolved question. 
In early work, \citet{oakes1977asymptotic} and \citet{efron1977efficiency} conducted information calculations assuming
absolute continuity of failure. 
In celebrated work, \citet{begun1983information} established that the Cox estimator attains the semiparametric efficiency bound 
assuming both failure and censoring are absolutely continuous. 
\citet{murphy2000profile} gave an alternative proof of efficiency using profile likelihood techniques, again assuming that
failure is absolutely continuous. 
\citet{hirose2011asymptotic} subsequently gave a direct profile-likelihood proof of the efficiency of the partial likelihood 
estimator in the continuous-failure setting, without imposing an analogous continuity assumption on the censoring distribution.

In practice, failure times may be tied, with two or more individuals experiencing the event at the same time. 
Acknowledging
that the observed failure distribution has mass points requires modifying the commonly used continuous-time Cox model. 
In his original work, \citet{cox1972regression}
both proposed a model that allows ties, which we call the Cox model throughout, and suggested a corresponding estimator,
which we call the Cox estimator throughout. 
Although ties are common in real data sets, the lion's share of theory is for the absolutely continuous case. 
Indeed, as far as the author knows, consistency, asymptotic normality, and semiparametric efficiency of the Cox 
estimator have not been established in the Cox model without assuming absolute continuity or assuming discreteness. 
When there are a small proportion of ties, the choice of how the estimator handles the ties is not generally consequential, whereas 
appreciable mass at some failure times can materially affect the resulting estimates. 

In addition to the Cox model, which has a logistic form at mass points, there are many alternative models such 
as a complementary log-log model at mass points, which results from grouping a continuous time Cox model 
\citep{kalbfleisch1973marginal,prentice1978regression} 
and a model allowing both continuous failure times and mass points while retaining the 
proportional hazards structure of the continuous-time Cox model \citep{prentice2003mixed}. 
Several estimators have been proposed under different models or as approximations, including those of 
\citet{peto1972ties}, \citet{breslow1974covariance}, and \citet{efron1977efficiency}. 
Nevertheless, large-sample analysis under models with mass points remains limited, with notable exceptions including 
\citet{prentice2003mixed}, who established
consistency and asymptotic normality for the \citet{breslow1974covariance} estimator in their proposed model, although they 
did not study efficiency. 
More recently, in very interesting work, \citet{cho2025accurate} proposed a Poisson--binomial approach to partial likelihood 
computation and established consistency and asymptotic normality under a grouped continuous failure-time model in which the grouping 
width shrinks with sample size, in the setting with grouping-induced ties, and they did not study the semiparametric efficiency bound.
In this work, we make three contributions. 
First, we derive the efficient influence function for the Cox 
coefficient when the failure distribution may be continuous, discrete, or mixed. 
Second, we establish that Cox’s exact partial-likelihood estimator is consistent, 
asymptotically normal, and semiparametrically efficient in the Cox model, 
with population mass points confined 
to a finite but unknown set. 
Third, we construct a profiled efficient score 
estimator with the same first-order asymptotic behavior and lower computational complexity per score evaluation,
based on a general construction of asymptotically efficient estimators, under the same finite set of population mass points assumption. 
We compare these estimators in simulations and illustrate them using data from advanced gastric-cancer trials. 
The remainder of the paper formulates the model and reviews existing estimators in \Cref{sec:back}, develops the 
efficiency theory and estimators in \Cref{sec:eff-score,sec:eff-est}, and presents the simulation and data analyses 
in \Cref{sec:sim,sec:data-example}.

\section{Background}
\label{sec:back}

In this section, we first present the observed data and Cox model, 
then give their latent-variable interpretation and review existing estimators for the Cox coefficient in the Cox model. 

\subsection{The Cox model}
\label{sec:back:mod}

We now introduce some familiar notation. Define the time at-risk $X>0$ and the failure indicator $\Delta \in \{0,1\}$.
Assume that $X \leq \tau$ for some $\tau < \infty$. Additionally define the baseline variables $\bL \in \R^p$ (for $p \geq
1$). Let $\P$ denote the distribution of $(\bL, X, \Delta)$. We also consider the derived stochastic processes
\begin{equation*}
    N_T(t) := I(\Delta=1, X \leq t), \hspace{5mm}
    N_C(t) := I(\Delta=0, X \leq t),
\end{equation*}
which provide the observed failure and censoring processes, respectively. 

Define the at-risk processes
\begin{equation*}
    Y(u) := I\{ N_C(u-)=0, N_T(u-)=0\}, \hspace{5mm}
    Y^{\dagger}(u) := I\{ N_C(u-)=0, N_T(u)=0 \}.
\end{equation*}
The at-risk process $Y = I(X \geq \cdot)$ is standard, and the modified at-risk process $Y^{\dagger}$ was recently exploited
by \citet{baer2023recurrent} while studying efficiency theory in survival analysis. We use these at-risk processes to
construct the functions
\begin{equation*}
    \Lambda_T^{\sharp}(t; \bl) 
    := \int_{(0, t]} \frac{\d \E \{ N_T(u) \mid \bL=\bl \} }{\E\{ Y(u) \mid \bL=\bl \}}, \hspace{5mm}
    \Lambda_C^{\dagger}(t; \bl) 
    := \int_{(0, t]} \frac{\d \E \{ N_C(u) \mid \bL=\bl \} }{\E\{ Y^{\dagger}(u) \mid \bL=\bl \}}, 
\end{equation*}
over $t > 0$. We analogously define $\Lambda_T^{\sharp}(\cdot)$ and $\Lambda_C^{\dagger}(\cdot)$ to be the corresponding
unconditional expressions. 
In the next section, \cref{sec:back:latent}, we discuss how these functions are related to the cumulative hazards for
failure and censoring, respectively.

We now define the Cox model which separates the longitudinal behavior and the effect of the baseline variables $\bL$ on
$\Lambda_T^{\sharp}$. For convenience we treat $\bL=\bzero$ as a supported covariate. Denote $\delta$ as the increment operation so
that $\delta F(t) := F(t) - F(t-)$ for a function $F$. It will be important throughout to keep in mind the contrast and
relationship between the differential $\d$ and the increment $\delta$. 
\begin{defn}
\label{defn:cox}
    The Cox model $\M$ is the set of finite variance distributions $\P$ for which there exists $\bbeta \in \R^p$ satisfying 
    \begin{equation*}
         \int_{(0,t]} \frac{\d \Lambda_T^{\sharp}(u; \bl)}{1-\delta \Lambda_T^{\sharp}(u; \bl)} = e^{\bbeta^{\top} \bl}
         \int_{(0,t]} \frac{\d \Lambda_T^{\sharp}(u; \bzero)}{1-\delta \Lambda_T^{\sharp}(u; \bzero)}
    \end{equation*}
    for all $t>0$ and almost all $\bl$. 
\end{defn}

The \emph{Cox coefficient} $\bbeta$ is the estimand, or parameter of interest. Under the technical assumptions in Section C.1 of the Supplementary Material
used in the theoretical development, the coefficient is unique and $\Lambda_T^{\sharp}(\cdot; \bzero)$ is well defined. 
Below are two examples of the model in simple cases. 

\begin{contexmp}[Continuous]
    Suppose $\Lambda_T^\sharp(\cdot; \bl)$ is absolutely continuous and hence dominated by the Lebesgue measure. Then 
    $\delta \Lambda_T^{\sharp}(t; \bl) = 0$ and $\Lambda_T^{\sharp}(t; \bl) := \int_{(0,t]} \lambda_T^{\sharp}(u; \bl) \,\d u$ 
    for the Lebesgue hazard $\lambda_T^{\sharp}$. Thus the Cox model $\M$ is defined by the familiar condition 
    \begin{equation*}
        \lambda_T^{\sharp}(t; \bl)
        = e^{\bbeta^{\top} \bl} \lambda_T^{\sharp}(t; \bzero), 
    \end{equation*}
    in which the hazards are proportional. 
\end{contexmp}

\begin{discexmp}[Discrete]
    Suppose that $\Lambda_T^\sharp(\cdot; \bl)$ is purely discrete and hence dominated by the counting measure. Then the Cox model $\M$ is defined by the
    condition
    \begin{equation*}
        \frac{\delta \Lambda_T^{\sharp}(t; \bl)}{1-\delta \Lambda_T^{\sharp}(t; \bl)} = e^{\bbeta^{\top} \bl} \frac{\delta
        \Lambda_T^{\sharp}(t; \bzero)}{1-\delta \Lambda_T^{\sharp}(t; \bzero)}, 
    \end{equation*}
    in which the odds are proportional. 

    Note, it is not possible to assume that $\delta \Lambda_T^{\sharp}(t; \bl) = e^{\bbeta^{\top} \bl} \delta
    \Lambda_T^{\sharp}(t; \bzero)$ without severe restrictions on the magnitude of $\bbeta^{\top} \bl$ since the probability $\delta
    \Lambda_T^{\sharp}(\cdot; \bl) \leq 1$. 
\end{discexmp}

The model was introduced by \citet{cox1972regression} to encompass already existing models that impose a parametric form for
$\Lambda_T^{\sharp}(t; \bzero)$. For example, under an exponential distribution and independent censoring assumption, the
function $\Lambda_T^{\sharp}(t; \bzero)$ is linear with a one-dimensional parameter. Some later work in discrete time has
employed alternatives to the Cox model, such as \citet{kalbfleisch1973marginal} who consider a cloglog-type model.

\subsection{Latent variables}
\label{sec:back:latent}

In this section we introduce in earnest the failure and censoring terms. Define the potential failure time $T^* > 0$ and the
potential censoring time $C^* > 0$. Let $\P^*$ denote the distribution of $(\bL, T^*, C^*)$. 
Throughout latent expressions, i.e. functionals of $\P^*$, are written with asterisks in the superscript.
Define the conditional cumulative hazard functions
\begin{equation*}
    \Lambda_T^*(t; \bl) := \int_{(0, t]} \frac{\d \P^* ( T^* \leq u \mid \bL=\bl )}{\P^* ( T^* \geq u \mid \bL=\bl )}, \hspace{5mm}
    \Lambda_C^*(t; \bl) := \int_{(0, t]} \frac{\d \P^* ( C^* \leq u \mid \bL=\bl ) }{\P^* ( C^* \geq u \mid \bL=\bl )}. 
\end{equation*}

Suppose that the observed data distribution $\P$ is determined by the latent distribution $\P^*$ through the relationship
$X := T^* \wedge C^*$ and $\Delta := I(T^* \leq C^*)$. 
The following result highlights that under conditionally independent censoring $\Lambda_T^{\sharp}$ and
$\Lambda_C^{\dagger}$ identify the conditional cumulative hazards for failure and censoring, respectively. 
\begin{lem}
\label{lem:cumhaz-identif}
    If $T^* \ind C^* \mid \bL$ on $C^* < T^*$, then $\Lambda_T^* = \Lambda_T^{\sharp}$ and $\Lambda_C^* = \Lambda_C^{\dagger}$
    over the data support. 
\end{lem}
\noindent The result is mostly standard, and a detailed proof is in \citet{baer2023recurrent} where the notation ``on $C^* <
T^*$'' is explained; note that the typical condition $T^* \ind C^* \mid \bL$ is sufficient as well. The result illustrates the
utility of the modified at risk process $Y^{\dagger}$ in that it's used to identify the cumulative censoring hazard. 

\begin{rem}
    \cref{lem:cumhaz-identif} shows that distributions $\P \in \M$ in the Cox model satisfy
    \begin{equation*}
        \int_{(0,t]} \frac{\d \Lambda_T^{*}(u; \bl)}{1-\delta \Lambda_T^{*}(u; \bl)} = e^{\bbeta^{\top} \bl} \int_{(0,t]} \frac{\d
        \Lambda_T^{*}(u; \bzero)}{1-\delta \Lambda_T^{*}(u; \bzero)}
    \end{equation*}
    under an additional (untestable) assumption of conditionally independent censoring. This is the perspective that
    motivated the introduction of the model \citep{cox1972regression}. 
\end{rem}

In this work we henceforth avoid making latent variable assumptions and instead directly study observed data functionals. 
This approach supports conceptual clarity and is now mainstream in some causal inference literature
\citep[cf.][]{kennedy2022semiparametric}. 
This is helpful from an applied perspective since it makes the results rely on fewer assumptions, and it is also helpful from
a theoretical perspective since the functionals then match the data. 
Of course, an immediate corollary of any result can be formed by assuming conditionally independent censoring and applying
\Cref{lem:cumhaz-identif}.

\subsection{Estimator review}
\label{sec:back:review}

\citet{cox1972regression} primarily studied the Cox coefficient $\bbeta$ when the failure distribution is absolutely
continuous. He proposed the seminal estimator that solves the estimating function
\begin{equation}
    \int_{(0, \tau]} \bL - \frac{ \E \left\{ \bL e^{\bL^{\top}\bbeta} Y(u) \right\} }{ \E \left\{ e^{\bL^{\top}\bbeta} Y(u) \right\} }
    \,\d N_T(u), \label{eq:cox-ef}
\end{equation}
where the nuisance parameters in the ratio are estimated for each $u$ using their sample analogues. 

The Cox estimator was motivated by the following argument. Consider $n$ data points indexed by $i$. In view of the absolute
continuity assumption, we assume here that each observed failure time occurs only once, i.e. that $X_i = X_j$ implies $i=j$
when $\Delta_i=1$ and $\Delta_j = 1$. The partial likelihood
\begin{equation}
    \prod_{i \, : \, \Delta_i = 1} \frac{ \exp \left( \bL_i^{\top}\bbeta \right) }{ \sum_{j \,:\, X_j \geq X_i} \exp \left(
    \bL_j^{\top}\bbeta \right) }, \label{eq:cox-partial}
\end{equation}
    may be viewed as a product of probabilities for the (informal) event that the $i$th data point is a failure among all $j$
that
could have failed at that time. 
The estimating function in \cref{eq:cox-ef} is the derivative of the log of \cref{eq:cox-partial}, after replacing its expectations 
with their sample analogues.

The matter of how to appropriately handle tied observed failure times has been debated since the introduction of the Cox model.
In Section 6 of his original paper, \citet{cox1972regression} discussed the issue and commented that ties are likely to be
observed in practice yet are not practically important when there are few. 
When there are an appreciable number of ties, \citet{cox1972regression} proposed replacing the partial likelihood in
\cref{eq:cox-partial} with 
\begin{equation}
     \prod_{i \in \mathcal{I}}
     \frac{ \exp\left( \sum_{j \in \mathcal{T}(i)} \bL_j^{\top}\bbeta \right) }
     { \sum_{\mathcal{S} \subseteq \mathcal{R}(i) \, : \, \card{\mathcal{S}} = \card{\mathcal{T}(i)}}
     \exp\left( \sum_{j \in \mathcal{S}} \bL_j^{\top}\bbeta \right) }, \label{eq:cox-partial2}
\end{equation}
where $\mathcal I$ contains one index for each distinct observed failure time, $\mathcal{R}(i) := \{ j \, : \, X_j \geq X_i \}$
is the usual risk set, $\mathcal{T}(i) := \{j \, : \, X_j = X_i, \Delta_j = 1\}$ is the set of tied failure data points,
and $\card{\cdot}$ denotes cardinality. 
The denominator is the sum over all possible combinations of data points with the same cardinality as $\mathcal{T}(i)$ that
could have failed. We may readily see that \cref{eq:cox-partial2} reduces to the probability in \cref{eq:cox-partial} when
there are no ties and hence $\mathcal{T}(i)$ is a singleton. In general, the denominator is a sum over $\card{\mathcal{R}(i)}$
choose $\card{\mathcal{T}(i)}$ terms, which quickly becomes computationally intractable; see \Cref{sec:meth:comp} for a more 
detailed analysis. 
We call the estimator that maximizes \Cref{eq:cox-partial2} the Cox estimator, although we'll use the term \emph{exact Cox estimator}
if any disambiguation may be helpful. 

There have been several other estimators for the Cox coefficient based on modifying the exact partial likelihood in
\Cref{eq:cox-partial2}. 
In the discussion of Cox's paper, \citet{peto1972ties} proposed the real probability and the rough probability to replace
the terms in the partial likelihood product. 
\citet{breslow1974covariance} suggested the \emph{Breslow approximation} that the probabilities in \cref{eq:cox-partial} be
used as if the data had no ties, even when there are ties, remarking that ``numerical comparisons \dots indicate that there
may be little practical difference \dots unless data are more heavily tied than here''. 
Finally, \citet{efron1977efficiency} proposed an approximation that was largely based on computational convenience. It is
now the default approach for dealing with ties in the \texttt{survival} package in \texttt{R} \citep{therneau2000cox,r2022},
and we compare against it in later simulations.

Taking a step back, \citet{cox1972regression} expressed that determining the efficiency of estimators in the Cox model was a
``major outstanding problem''. He suggested that the Cox estimator should be asymptotically efficient since ``no information
can be contributed about $\bbeta$ by time intervals in which no failures [are observed] because the component of [the
baseline hazard] might conceivably be identically zero in such intervals.'' 
In absolutely continuous time, \citet{oakes1977asymptotic, efron1977efficiency} study efficiency of the Cox estimator and
\citet{begun1983information} fully established the efficiency of the Cox estimator in the Cox model using the now seminal
approach of Stein \citep{stein1956efficient,vaart2021stein}. 
To our knowledge, the semiparametric efficiency bound and efficiency of Cox’s exact estimator have not been studied in the 
presence of finitely many failure mass points without the distribution being fully discrete. Existing alternatives have primarily 
been motivated by model choice, robustness, or computational considerations rather than attainment of this bound.

\section{The efficiency bound}
\label{sec:eff-score}

In this section, we develop the class of influence functions and the efficient score for $\bbeta$ in the Cox model $\M$.

\subsection{The density}

The first step is to write down the density of the observed data. Because we simultaneously consider cases in which the
distribution $\P$ is continuous, discrete, singular, or mixed, this step is somewhat involved. 

The density of a distribution is defined with respect to a dominating measure as a Radon-Nikodym derivative
\citep{durrett2019probability}. For example, if $\mu_{\bL}$ denotes the distribution of the baseline variables $\bL$ and $\nu_{\bL}$
denotes a dominating measure (such as the Lebesgue measure), then its density (or probability density function) is
$\frac{\d \mu_{\bL}}{\d \nu_{\bL}}$. Throughout we suppress the ``denominator'' of the density following
\citet{andersen1993statistical}. This convention is helpful since the dominating measure will often be a mixture of the
Lebesgue and counting measures.

\subsubsection{The general density}

We now state the density of the observed data $\bL,X,\Delta$ without any extra assumptions. Examples for the continuous and
discrete cases follow. 
For a function $F(\cdot)$, denote its product integral as
\begin{equation*}
    \Prodi_{u \in (0,t]} \left\{ 1 - \d F(u) \right\}.
\end{equation*}
The product integral is thoroughly studied by \citet{gill1990survey} and satisfies important properties such as
multiplicativity. Now, the density follows from Jacod's 
formula for likelihood ratios 
\citep[Theorem II.7.2]{andersen1993statistical}. 

\begin{prop}
\label{prop:density}
    The density $\d \P (\bl,x,d)$ of the observed data $(\bL,X,\Delta)$, for $d\in\{0,1\}$, is
    \begin{align*}
        \d \mu_{\bL} (\bl) 
        \left\{ \d \Lambda_T^{\sharp}(x; \bl) \right\}^{d} 
        \left\{ \d \Lambda_C^{\dagger}(x; \bl) \right\}^{1-d} 
        \frac{\Prodi_{u \in (0, x]} 1 - \d \Lambda_T^{\sharp}(u; \bl)}{\left\{ 1 - \delta \Lambda_T^{\sharp}(x; \bl)
        \right\}^{d}} 
        \Prodi_{u \in (0, x)} 1 - \d \Lambda_C^{\dagger}(u; \bl). 
    \end{align*}
\end{prop}

This notation for the density in \cref{prop:density} can seem foreign, so below we give two simple examples.

\begin{contexmp}[Continuous]
\label{exmp:dens-cont}
    Suppose that all random variables other than $\Delta$ are absolutely continuous and hence dominated by the Lebesgue measure. Then $\d \mu_{\bL}$
    is the Lebesgue density, $\d \Lambda_T^{\sharp}$ and $\d \Lambda_C^{\dagger}$ are Lebesgue hazards, and $\delta
    \Lambda_T^{\sharp}(x; \bl), \delta \Lambda_C^{\dagger}(x; \bl) = 0$. The product integrals in \Cref{prop:density} reduce to
    exponential survival terms, so the density is
    \begin{align*}
        \d \mu_{\bL} (\bl)
        \left\{ \d \Lambda_T^{\sharp}(x; \bl) \right\}^{d}
        \left\{ \d \Lambda_C^{\dagger}(x; \bl) \right\}^{1-d}
        \exp \left\{ - \Lambda_T^{\sharp}(x; \bl) - \Lambda_C^{\dagger}(x; \bl) \right\},
    \end{align*}
    which is the familiar continuous-time density. The product-integral reduction is given in
    Section A.2 of the Supplementary Material.
\end{contexmp}

\begin{discexmp}[Discrete]
\label{exmp:dens-discr}
    Suppose that the failure and censoring distributions are discrete and hence dominated by the counting measure. Then $\d
    \Lambda_T^{\sharp}(t; \bl) = \delta \Lambda_T^{\sharp}(t; \bl)$ and $\d \Lambda_C^{\dagger}(t; \bl) = \delta
    \Lambda_C^{\dagger}(t; \bl)$ are counting hazards. The product integrals in \Cref{prop:density} reduce to ordinary
    products, so the density is
    \begin{align*}
        & \d \mu_{\bL} (\bl)
        \left\{ \delta \Lambda_T^{\sharp}(x; \bl) \right\}^{d}
        \left\{ \delta \Lambda_C^{\dagger}(x; \bl) \right\}^{1-d}
        \frac{\prod_{0 < u \leq x} \left\{ 1 - \delta \Lambda_T^{\sharp}(u; \bl) \right\}}{\left\{ 1 - \delta
        \Lambda_T^{\sharp}(x; \bl) \right\}^{d}}
        \prod_{0 < u < x} \left\{ 1 - \delta \Lambda_C^{\dagger}(u; \bl) \right\},
    \end{align*}
    where the products are over the finite number of fixed possible jump points. The reduction and a direct probability
    calculation are given in Section A.3 of the Supplementary Material.
\end{discexmp}

Now, we state the density under the Cox model, in which the
identified cumulative hazard $\Lambda_T^{\sharp}$ for failure satisfies
\begin{equation}
    \Lambda_T^{\sharp}(t; \bL) = \int_{(0,t]} \frac{e^{\bbeta^{\top} \bL} \d\Lambda_T^{\sharp}(u; \bzero)}{1 +
    (e^{\bbeta^{\top} \bL} - 1) \delta\Lambda_T^{\sharp}(u; \bzero)}. \label{eq:lam-t-cox}
\end{equation}
The density is below, parameterized by $\d \mu_{\bL}$, $\Lambda_C^{\dagger}(\cdot; \bL)$,
$\Lambda_T^{\sharp}(\cdot; \bzero)$, and $\bbeta$.

\begin{cor}
\label{cor:dens}
    The density $\d \P (\bl,x,d)$ of the observed data $(\bL,X,\Delta)$, for $d\in\{0,1\}$, in the Cox model $\M$ is
    \begin{align*}
        & \d \mu_{\bL} (\bl) 
        \left\{ \frac{e^{\bbeta^{\top} \bl} \d \Lambda_T^{\sharp}(x; \bzero)}{1 - \delta \Lambda_T^{\sharp}(x; \bzero)} \right\}^{d}
        \left\{ \d \Lambda_C^{\dagger}(x; \bl) \right\}^{1-d} \\
        & \hspace{20mm} \times \Prodi_{u \in (0, x]} \left\{ 1 - \frac{e^{\bbeta^{\top} \bl} \d
        \Lambda_T^{\sharp}(u; \bzero)}{1 + (e^{\bbeta^{\top} \bl} - 1) \delta \Lambda_T^{\sharp}(u; \bzero)} \right\}
        \Prodi_{u \in (0, x)} \left\{ 1 - \d \Lambda_C^{\dagger}(u; \bl) \right\}
    \end{align*}
\end{cor}

For the tangent space calculations, we use the alternative cumulative hazards $\Lambda_T^{\dagger}$ and
$\Lambda_C^{\ddagger}$, defined in Section B.1 of the Supplementary Material, which admit convenient multiplicative tilts.

\subsection{The class of influence functions} 

The section culminates with a theorem that furnishes all estimating functions for $\bbeta$ in the Cox model $\M$. A
proposition providing the nuisance tangent space sets the stage. 

Define the centered stochastic processes 
\begin{equation*}
    M_T^{\sharp}(t; \bL)
    := N_T(t) - \int_{(0,t]} Y(u) \,\d \Lambda_T^{\sharp}(u; \bL), \hspace{5mm} 
    M_C^{\dagger}(t; \bL)
    := N_C(t) - \int_{(0,t]} Y^{\dagger}(u) \,\d \Lambda_C^{\dagger}(u; \bL).
\end{equation*}
The process $M_T^{\sharp}$ is a well-studied Doob-Meyer martingale, while the process $M_C^{\dagger}$ is thoroughly studied in
\citet{baer2023censoring}. 
Consider the vector space $L_2^0(\P)$ of transformations of the observed data $(\bL,X,\Delta)$ with the same dimension as
$\bbeta$ and with finite variance and expectation zero. The nuisance tangent space is a subspace generated by scores with respect to the nuisance
parameters $\d \mu_{\bL}$, $\Lambda_C^{\dagger}(\cdot; \bL)$, and $\Lambda_T^{\sharp}(\cdot)$. 

\begin{prop}
\label{prop:nuis-tangentspace}
    The nuisance tangent space for $\bbeta$ at $\P$ in the Cox model $\M$ is spanned by 
    \begin{equation*}
        \balpha_1(\bL) 
        + \int_{(0, \tau]} \balpha_2(u; \bL) \,\d M_C^{\dagger}(u; \bL)
        + \int_{(0, \tau]} \balpha_3(u) \,\d M_T^{\sharp}(u; \bL),
    \end{equation*}
    where $\balpha_1,\balpha_2,\balpha_3$ are functions with the same dimension as $\bbeta$ and $\E\{\balpha_1(\bL)\}=0$. 
\end{prop}
\noindent The proof finds the tangent spaces in $\M$ for each of the nuisance parameters $\d \mu_{\bL}$,
$\Lambda_C^{\dagger}(\cdot; \bL)$, and $\Lambda_T^{\sharp}(\cdot)$ separately. Then, since each of the nuisance parameters
is variation independent, the nuisance tangent space is their direct sum. 
The functions $\balpha_1,\balpha_2,\balpha_3$ are also constrained so that each element displayed in \Cref{prop:nuis-tangentspace} 
has finite variance, although we omit this specification in later development.

Recall that an influence function is the linearization of an (asymptotically linear) estimator and characterizes the
first-order asymptotic behavior of the estimator. The nuisance tangent space plays a central role in semiparametric
efficiency theory since all influence functions of semiparametric estimators are orthogonal to the nuisance tangent space
\citep[Section 1.4.2]{van2003unified}. The next result characterizes this space.

\begin{thm}
\label{thm:orth-comp}
    The orthogonal complement of the nuisance tangent space for $\bbeta$ at $\P$ in $\M$ is spanned by
    \begin{equation*}
        \int_{(0, \tau]} \balpha(u; \bL)-\frac{\E \left[ \balpha(u; \bL) \left\{ 1 - \delta \Lambda_T^{\sharp}(u; \bL) \right\}^{2}
        Y(u) e^{\bbeta^{\top}\bL} \right]}{\E \left[ \left\{ 1 - \delta \Lambda_T^{\sharp}(u; \bL) \right\}^{2} Y(u)
        e^{\bbeta^{\top}\bL} \right]} \,\d M_T^{\sharp}(u; \bL), 
    \end{equation*}
    for a function $\balpha(\cdot; \bL)$ with the same dimension as $\bbeta$. 
\end{thm}

The functionals in the orthogonal complement are generally variation dependent with $\bbeta$. Using \cref{eq:lam-t-cox}, a
typical element of the orthogonal complement may be rewritten as 
\begin{equation}
    \int_{(0,\tau]} \balpha(u; \bL) - \frac{\E \left[ \frac{Y(u) e^{\bbeta^{\top} \bL}}{\{ 1 + (e^{\bbeta^{\top} \bL} - 1)
    \delta\Lambda_T^{\sharp}(u; \bzero) \}^2} \balpha(u; \bL) \right]}{\E \left[ \frac{Y(u) e^{\bbeta^{\top} \bL}}{\{ 1 + (e^{\bbeta^{\top}
    \bL} - 1) \delta\Lambda_T^{\sharp}(u; \bzero) \}^2} \right]} \left\{ \d N_T(u) - Y(u) \frac{e^{\bbeta^{\top} \bL}
    \d\Lambda_T^{\sharp}(u; \bzero)}{1 + (e^{\bbeta^{\top} \bL} - 1) \delta\Lambda_T^{\sharp}(u; \bzero)} \right\}.
    \label{eq:orth-comp-varind}
\end{equation}
This representation shows the only nuisance parameter is $\Lambda_T^{\sharp}(\cdot; \bzero)$, which is variation independent of
$\bbeta$.

Every influence function for $\bbeta$ must lie in the orthogonal complement
of the nuisance tangent space and satisfy the usual normalization condition
with respect to the score for $\bbeta$
\citep[Lemma 1.3]{van2003unified}. Thus \cref{thm:orth-comp} supplies the class
of influence functions; imposing the normalization condition
identifies the influence functions within this class.
In turn, an influence function may be used as an estimating function; under
conditions on nuisance parameter estimation, the resulting estimator has
that a related influence function \citep{van1998asymptotic}. 
Therefore, under regularity conditions, the
orthogonal complement encompasses all estimating functions for $\bbeta$, up to asymptotic equivalence, and
\cref{thm:orth-comp} thus characterizes all semiparametric estimators in $\M$ for $\bbeta$, up to asymptotic equivalence.

\subsection{The efficient influence function}

Although \cref{thm:orth-comp} identifies the class of candidate influence
functions for estimating the Cox coefficient, it does not determine which one attains the efficiency bound. In
this section, we derive the efficient influence function, namely the
normalized member of this class having the smallest covariance matrix. Later, this is used to construct an 
asymptotically efficient estimator.

\begin{thm}
\label{thm:eff-score}
    
    The efficient influence function for $\bbeta$ at $\P$ in $\M$ is $\left\{ \E \left( \bS_{\mathrm{eff}}
    \bS_{\mathrm{eff}}^{\top} \right) \right\}^{-1} \bS_{\mathrm{eff}}$, where $\bS_{\mathrm{eff}}$ is the expression in
    \cref{thm:orth-comp} when $\balpha(u; \bL) = \bL$, that is 
    \begin{equation*}
        \int_{(0, \tau]} \bL - \frac{\E \left[ \bL \left\{ 1 - \delta \Lambda_T^{\sharp}(u; \bL) \right\}^{2}
        Y(u) e^{\bbeta^{\top} \bL} \right]}{\E \left[ \left\{ 1 - \delta \Lambda_T^{\sharp}(u; \bL) \right\}^{2} Y(u)
        e^{\bbeta^{\top}\bL} \right]} \,\d M_T^{\sharp}(u; \bL), 
    \end{equation*}
\end{thm}

\begin{discexmp}[Discrete]
\label{exmp:eff-score-discrete}
    Consider again that
    \begin{align*}
        \d \Lambda_T^{\sharp}(t; \bl)
        & = \delta \Lambda_T^{\sharp}(t; \bl)
        = \P ( X = t, \Delta = 1 \mid \bL=\bl, X \geq t )
    \end{align*}
    are counting hazards. Under the Cox model in \Cref{defn:cox}, the discrete hazard satisfies
    \begin{equation*}
        \logit\{ \delta \Lambda_T^{\sharp}(t; \bL) \}
        = \bbeta^{\top} \bL + \logit \{ \delta \Lambda_T^{\sharp}(t; \bzero) \}.
    \end{equation*}
    Thus the contribution of the failure process to the likelihood is a weighted logistic regression with nuisance parameter
    $\alpha_t := \logit \{ \delta \Lambda_T^{\sharp}(t; \bzero) \}$. Projecting the score for $\bbeta$ off the scores for
    $(\alpha_t)$ gives
    \begin{align*}
        \bS_{\mathrm{eff}}
        = \int_{(0,\tau]} \left\{ \bL - \frac{\E\left[
        Y(t) e^{\bbeta^{\top}\bL} \{1-\delta\Lambda_T^{\sharp}(t;\bL)\}^2 \, \bL
        \right]}{\E\Big[ Y(t)e^{\bbeta^{\top}\bL}\{1-\delta\Lambda_T^{\sharp}(t;\bL)\}^2
        \Big]} \right\} \d M_T^{\sharp}(t; \bL), 
    \end{align*}
    using standard parameteric likelihood theory, where the integral is over the discrete jump times. 
    The likelihood factorization and score projection calculation are given in Section B.5 of the Supplementary Material. 
\end{discexmp}

\begin{contexmp}[Continuous]
    Suppose $\Lambda_T^\sharp(\cdot; \bl)$ is absolutely continuous and hence dominated by the Lebesgue measure. 
    Then $\delta \Lambda_T^{\sharp}(t;
    \bl) = 0$ for all $t>0$ and supported $\bl$ and $\Lambda_T^{\sharp}(t; \bl) := \int_{(0,t]} \lambda_T^{\sharp}(u; \bl) \,\d u$
    for the Lebesgue hazard $\lambda_T^{\sharp}$. 
    In this case we readily see that
    \begin{equation*}
        \bS_{\mathrm{eff}}
        = \int_{(0,\tau]} \bL - \frac{\E \left\{ \bL \, e^{\bbeta^{\top} \bL} Y(u) \right\}}{\E \left\{ e^{\bbeta^{\top} \bL} Y(u)
        \right\}} \,\d M_T^{\sharp}(u; \bL),
    \end{equation*}
    a familiar form that appears in e.g. \citet[Eq. 3.25]{fleming1991counting} and \citet[Page
    125]{tsiatis2006semiparametric}. 

\end{contexmp}

\section{Efficient estimation}
\label{sec:eff-est}

Consider $n$ observations drawn independently from a true distribution $\P_0 \in \M$ in the Cox model. Denote the empirical
distribution by $\P_n$ and the corresponding expectation by $\E_n$.
Denote all functionals evaluated at the true distribution $\P_0$ as subscripted by $0$, so that e.g. $\bbeta_0$ is the
estimand.

Define $\mJ(\P) = \left\{ t\in(0,\tau] : \P\{\delta N_T(t)>0\}>0 \right\}$ as the set of population failure-time mass points (or jumps). 
Throughout this section we make the following assumption. 
\begin{asp}
\label{asp:jumps}
    $\mJ(\P_0)$ is finite. 
\end{asp}
\noindent Restricting the Cox model to distributions satisfying \Cref{asp:jumps} does not change the efficiency bound at $\P_0$, 
since the submodels generating the tangent space in \Cref{prop:nuis-tangentspace} remain available under this restriction. 

In the Cox model, the set $\mJ(\P) = \left\{ t\in(0,\tau] : \delta\Lambda_T^\sharp(t;\bzero)>0 \right\}$ can be expressed 
in terms of the baseline cumulative hazard, and 
the time points supporting ties are common across covariate values.

\subsection{The Cox estimator}
\label{sec:eff-est:cox}

In this section we establish that the Cox estimator $\widehat{\bbeta}_{\text{Cox}}$ that maximizes the partial
likelihood in \Cref{eq:cox-partial2} is consistent for $\bbeta_0$ and asymptotically efficient for the Cox coefficient. 

\begin{thm}
\label{thm:exact}
    Assume the technical conditions in Section C.1 of the Supplementary Material. 
    If $\M$ is correctly specified and \Cref{asp:jumps} holds, then $\widehat{\bbeta}_{\text{Cox}}$ is consistent and has influence function
    $\left\{ \E \left( \bS_{\mathrm{eff}} \bS_{\mathrm{eff}}^{\top} \right) \right\}^{-1} \bS_{\mathrm{eff}}$. 
\end{thm}

\noindent The proof critically uses the results in \citet{arratia2005local}, which studies the asymptotic behavior of
conditional logistic regression in matched case-control study designs using the rejective sampling of \citet{hajek1964asymptotic}.

The result shows that the Cox estimator is 
consistent, extending the result first established by \citet{tsiatis1981large}, and 
asymptotically efficient, extending the result first established by \citet{begun1983information}
and later extended by \citet{murphy2000profile,hirose2011asymptotic},  
to allow for discontinuities in the failure and censoring distributions and hence be applicable to the empirically common
scenario of there being ties. 
The theorem also shows that suitably constructed Wald confidence intervals asymptotically achieve their nominal coverage,
making rigorous inference possible with the Cox estimator when there are ties.

\subsection{General coefficient estimation}
\label{sec:eff-est:coef}

In this section we develop an estimator $\widehat{\bbeta}_n$ for $\bbeta_0$ whose influence function is the efficient influence function
in \cref{thm:eff-score}. 

Define the estimator $\widehat{\bbeta}_n$ as solving the estimating equation associated with 
using the efficient score $\bS_{\mathrm{eff}}$ as the estimating function given by 
\begin{equation}
    \int_{(0,\tau]} \left[ \bL - \frac{\E \left\{ \frac{e^{\bbeta^{\top} \bL}}{\{ 1 + (e^{\bbeta^{\top} \bL} - 1)
    \delta\Lambda_T^{\sharp}(u; \bzero) \}^2} Y(u) \bL \right\}}{\E \left\{ \frac{e^{\bbeta^{\top} \bL}}{\{ 1 + (e^{\bbeta^{\top} \bL} -
    1) \delta\Lambda_T^{\sharp}(u; \bzero) \}^2} Y(u) \right\}} \right] \left\{ \d N_T(u) - Y(u) \frac{e^{\bbeta^{\top} \bL}
    \d\Lambda_T^{\sharp}(u; \bzero)}{1 + (e^{\bbeta^{\top} \bL} - 1) \delta\Lambda_T^{\sharp}(u; \bzero)} \right\},
    \label{eq:eff-est-function}
\end{equation}
which has three nuisance parameters. The two nuisance parameters forming the numerator and denominator of the ratio may 
be estimated using their sample analogues, while the baseline cumulative hazard $\Lambda_T^{\sharp}(\cdot; \bzero)$ is the subject
of the assumptions below. 

\begin{asp}
\label{asp:incr-rate}
    $\sup_{t \in (0,\tau]} \left| \widehat{\Lambda}^{\sharp}_{T,n}(t; \bzero) - \Lambda^{\sharp}_{T,0}(t; \bzero) \right|
    = o_{\P_0}(1)$.
\end{asp}

\begin{asp}
\label{asp:incr-rate2}
    $\left[ \sum_{u \in (0,\tau]} \left\{ \delta \widehat{\Lambda}^{\sharp}_{T,n}(u; \bzero) - \delta \Lambda^{\sharp}_{T,0}(u; \bzero)
    \right\}^2 \right]^{1/2}
    = o_{\P_0}(n^{-1/4})$. 
\end{asp}

\begin{thm}
\label{thm:estimator}
    If \Cref{asp:jumps,asp:incr-rate,asp:incr-rate2} and the regularity conditions in Section C.1 of the Supplementary Material hold, then $\widehat{\bbeta}_n$ is
    locally efficient at $\P_0$ in $\M$. 
\end{thm}

When the theorem holds, the estimator satisfies the asymptotic expansion
\begin{equation*}
    \widehat{\bbeta}_n - \bbeta_0
    = \left[ \E_0 \left\{  \bS_{\text{eff}}(\mathcal{O}) \bS_{\text{eff}}(\mathcal{O})^{\top} \right\} \right]^{-1} 
     \E_n \left\{ S_{\text{eff}}(\mathcal{O}) \right\} + o_{\P_0}(n^{-1/2}). 
\end{equation*}
We use this expansion to obtain large-sample Wald standard errors, replacing the leading expectation with its empirical
analog.

\subsection{A proposed estimator}

We now introduce a proposed estimator based on a proposed estimator of the baseline cumulative hazard. 
Let $\mJ := \{ t \in (0,\tau] \, : \, \P\{ \delta N_T(t) > 0 \} > 0 \}$ denote the failure jumps. Consider an
estimator $\widehat{\mJ}$ that converges to $\mJ$ as $n \to \infty$; throughout we put 
$\widehat{\mJ} := \left\{ t\in(0,\tau]: \sum_{i=1}^n \delta N_{T,i}(t) > 1 \right\}$. 
For each $\bbeta$, define $\widetilde{\Lambda}_{T,n}^{\sharp}(\cdot;\bzero, \bbeta)$ as piecewise constant with increments
$\delta\widetilde{\Lambda}_{T,n}^{\sharp}(t;\bzero, \bbeta)$ satisfying the profiled estimating equation  
\begin{equation}
    \E_n \left\{ \delta N_T(t) - Y(t) \frac{ e^{\bbeta^{\top}\bL}\delta\widetilde{\Lambda}_{T,n}^{\sharp}(t;\bzero, \bbeta)}{
    1+\{e^{\bbeta^{\top}\bL}-1\}\delta\widetilde{\Lambda}_{T,n}^{\sharp}(t;\bzero, \bbeta) I( t \in \widehat{\mathcal{J}}) } \right\} = 0,
    \label{eq:profile-baseline}
\end{equation}
which ensures the fitted baseline hazard predicts the correct total number of failures across time. 
Then, define $\widetilde{\bbeta}_n$ as the solution of 
\begin{equation}
    \E_n\left[\int_{(0,\tau]}\bL \left\{ \d N_T(u) - Y(u) \frac{ e^{\bbeta^{\top}\bL}\d\widetilde{\Lambda}_{T,n}^{\sharp}(u;\bzero,
    \bbeta) }{ 1+\{e^{\bbeta^{\top}\bL}-1\}\delta\widetilde{\Lambda}_{T,n}^{\sharp}(u;\bzero, \bbeta) I(u \in \widehat{\mathcal{J}}) }
    \right\} \right] = 0, \label{eq:profile-beta}
\end{equation}
which is based on the efficient score display in \Cref{eq:eff-est-function} after noting that the term subtracted from $\bL$
must vanish due to \cref{eq:profile-baseline}.

\begin{cor}
\label{cor:profile-estimator}
    Assume the regularity conditions in Section C.1 of the Supplementary Material. If \Cref{asp:jumps} holds, 
    then $\widetilde{\bbeta}_n$ is locally efficient at $\P_0$ in $\M$. 
\end{cor}

When there are no ties and $\widehat{\mathcal{J}}=\{\}$, the proposed baseline cumulative hazard estimator
$\widetilde{\Lambda}_{T,n}^{\sharp}(\cdot; \bzero, \widetilde{\bbeta}_n)$ defined in \Cref{eq:profile-baseline} is simply the
estimator of \citet{breslow1972hazard},
\begin{equation*}
    \widehat{\Lambda}_{T,n,\text{Breslow}}^{\sharp}(t; \bzero)
    := \int_{(0,t]} \frac{\d \E_n \{ N_{T}(u) \}}{\E_n \{ Y(u) e^{\widehat{\bbeta}^{\top} \bL} \}},
\end{equation*}
where $\widetilde{\bbeta}_n = \widehat{\bbeta}_n$ is the Cox estimator. 
When $N_T$ is not absolutely continuous, recall from \cref{eq:lam-t-cox} that under the Cox model the cumulative hazard is 
\begin{equation*}
    \Lambda_T^{\sharp}(t; \bL) 
    = \int_{(0,t]} \frac{e^{\bbeta^{\top} \bL} \,\d \Lambda_T^{\sharp}(u; \bzero)}{1 + ( e^{\bbeta^{\top} \bL} - 1 ) \delta
    \Lambda_T^{\sharp}(u; \bzero)}, 
\end{equation*}
which modifies the form of the martingale increments that are set to zero in \cref{eq:profile-baseline}.

\subsection{Computational complexity}
\label{sec:meth:comp}

In this section, we compare the computational complexity of evaluating the estimating functions defining the Cox and
proposed estimators. We consider one evaluation at a fixed value of $\bbeta$, after the observations have been sorted by
time in $O(n\log n)$ complexity. 
When comparing orders in risk-set size, we hold $p$ fixed.

Recall that $\mathcal I$ contains one index for each distinct observed failure time, $\mathcal R(i):=\{j:X_j\ge X_i\}$
is the risk set at failure time $X_i$, and $\mathcal T(i):=\{j:X_j=X_i,\ \Delta_j=1\}$ is the set of individuals whose observed
failure occurs at the same time as the $i$th individual. 
\citet{howard1972discussion} developed a
dynamic-programming recursion that evaluates the exact partial likelihood in \cref{eq:cox-partial2} in 
$O\left[ np + \sum_{i\in\mathcal I} \card{\mathcal T(i)} \left\{\card{\mathcal R(i)}-\card{\mathcal T(i)}\right\} \right]$
operations; see also \citet{gail1981likelihood}. 
Since the Cox and proposed estimators are both defined through $p$-dimensional
estimating equations, we instead focus on comparing the costs of evaluating these equations.

The Howard--Gail recursion can also be used to evaluate the exact Cox score. After the observations have been sorted, 
the score can be evaluated in
\begin{equation}
    O\left[
        np + \sum_{i\in\mathcal I \, : \, X_i\in\widehat{\mathcal J}}
        \biggl\{
            \card{\mathcal T(i)} \Bigl(\card{\mathcal R(i)}-\card{\mathcal T(i)}\Bigr) + p\card{\mathcal R(i)}
        \biggr\} \right] \label{eq:exact-score-complexity}
\end{equation}
operations. 
The term \(np\) accounts for forming the linear predictors and evaluating the ordinary Cox-score contributions 
at untied failure times. At each tied failure time, the first term inside the summation is the cost of the Howard--Gail 
recursion, while the second is the cost of assembling the resulting \(p\)-dimensional score contribution. 

We next consider the proposed efficient score estimating equation in \cref{eq:profile-beta}, including the profiled baseline
increments in \cref{eq:profile-baseline}. Outside the set of ties $\widehat{\mathcal J}$, the profiled increment has the closed-form Breslow
solution, and the resulting score contribution is the ordinary Cox score. As above, all such contributions can be evaluated
together in $O(np)$ operations. At each failure time in $\widehat{\mathcal J}$, evaluating the scalar equation in
\cref{eq:profile-baseline} requires a sum over the risk set and hence $O\{\card{\mathcal R(i)}\}$ operations. Once its root
has been obtained, evaluating the corresponding $p$-dimensional contribution to \cref{eq:profile-beta} requires
$O\{p\card{\mathcal R(i)}\}$ operations. Since $p\geq 1$ and the scalar equation is solved to a fixed tolerance, the proposed
estimating equation can therefore be evaluated in 
\begin{equation}
    O\left( np+ p\sum_{i\in\mathcal I \, : \, X_i\in\widehat{\mathcal J}} \card{\mathcal R(i)} \right)
    \label{eq:proposed-score-complexity}
\end{equation}
operations.

The two scores share the cost of forming weighted $p$-dimensional sums over the risk sets. The exact Cox score additionally
requires the dynamic-programming calculation
\begin{equation*}
    O \left\{ \sum_{i\in\mathcal I \, : \, X_i\in\widehat{\mathcal J}}
    \card{\mathcal T(i)}
    \Bigl(\card{\mathcal R(i)}-\card{\mathcal T(i)}\Bigr) \right\}. 
\end{equation*}
When there are no ties, $\widehat{\mathcal J}=\{\}$ and both scores can be evaluated in $O(np)$ operations; indeed
the scores are the same. 
When the
sizes of the tied failure sets remain bounded, the two scores generally have the same computational order. 
The proposed efficient score estimating equation is asymptotically faster when both the numbers of tied failures and 
nonfailures in a risk set grow. 
In particular, if tied failures constitute a nondegenerate fraction of the risk set and \(p\) is fixed, the exact score and 
the efficient score estimating equation calculations are respectively quadratic and linear in the risk-set size.

\section{Simulation}
\label{sec:sim}

In this section, we study the empirical performance of the considered estimators as the discrete mass of the failure
distribution varies. 
We considered the Cox estimator, implemented using the ``exact'' option in coxph in the survival R package, and Efron's
approximation, implemented using the ``efron'' option in coxph in the same survival R package
\citep{cox1972regression,efron1977efficiency,therneau2023package}, along with the proposed estimator. 

\subsection{The setting} 

Throughout the baseline covariates $L_1 \sim \mathrm{Bernoulli}(1/2)$ and $L_2 \sim \mathcal{N}(0,1)$, independently. The
censoring hazard is 
\begin{equation*}
    \frac{\d \Lambda_C^{\dagger}(t; \bL)}{\d t} 
    = 0.1 \exp\Bigl\{ \log(2) L_1 - \log(2) L_2 \Bigr\}
\end{equation*}
for $t < \tau$, along with administrative censoring at $\tau=5$. 
For the failure distribution, define the mass points $0 < t_1 < \cdots < t_k \leq \tau$ for $k=3$. 
We take $t_j = \frac{j}{4} \tau$ for $j=1,2,3$. 
The failure cumulative
hazard $\Lambda_T^{\sharp}(\cdot; \bl)$ has jump discontinuities at $t_1, \dots, t_k$ and is continuous otherwise. 
Off the mass points, the failure hazard $\d\Lambda_T^{\sharp}(t;\bzero) / \d t = \frac{1-q}{\tau}$ is constant. 
Define $S_T^\sharp(t; \bzero) := \prodi_{u\in(0,t]} \left\{1-\d\Lambda_T^\sharp(u;\bzero)\right\}$ as the baseline survival
probability, and define 
\begin{equation*}
    q 
    := \frac{\sum_{j=1}^k S_T^\sharp(t_j-; \bzero) \delta\Lambda_T^\sharp(t_j;\bzero)}{1-S_T^\sharp(\tau; \bzero)} 
\end{equation*}
as the fraction of failures by $\tau$ that occur at mass points before $\tau$. When $q=0$ the failure distribution is
absolutely continuous, and when $q=1$ it is discrete. Then, solving for the hazard increment, we define the failure
cumulative hazard increment 
\begin{equation*}
    \delta\Lambda_T^\sharp(t_j;\bzero)
    := \frac{q\{1-S_T^\sharp(\tau; \bzero)\}}{k\,S_T^\sharp(t_j-; \bzero)}. 
\end{equation*}
Additionally, we define the estimand $\bbeta_0 := (\log(2),-\log(2))^{\top}$. 

Results are based on 1000 Monte Carlo samples in each setting and $n=500$.

\subsection{The output and interpretation}

\begin{figure}
    \includegraphics[scale=.85]{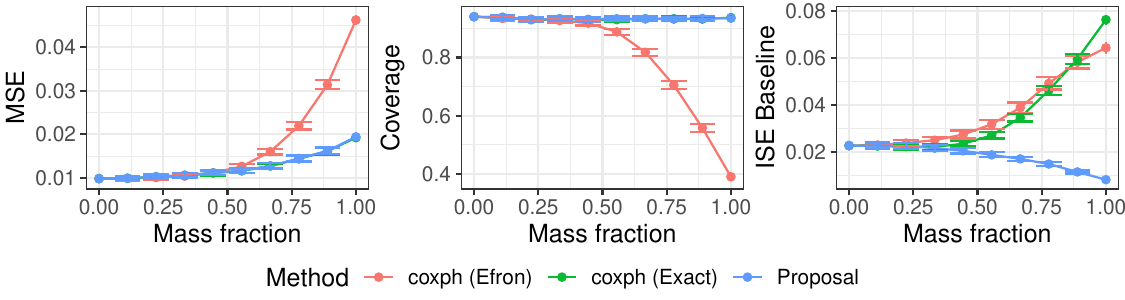}
    \caption{A comparison of the Cox (exact) estimator, the Efron approximation, and the proposed estimator in
    terms of the coefficient mean squared error (MSE), Wald confidence interval coverage, and integrated squared
    error (ISE) of the baseline cumulative hazard estimator as the fraction of the failure distribution with mass
    increases.}
\end{figure}

We consider ten equally spaced mass fractions $q$ spanning $[0,1]$, and the mean proportion censored ranged from $0.170$ to
$0.227$ across the ten settings. 

We start by considering the two left-most panes, which show the mean squared coefficient error (MSE) and the average Wald
coverage over the two coefficients. 
The proposed estimator (denoted ``Proposal'') and the Cox estimator (denoted ``Cox (exact)'') gave nearly identical results
throughout. What appears to be the single blue (Proposal) curve is actually both the blue and the green (Cox, exact) curves
on top of each other. 
The Efron approximation (denoted ``Cox (Efron)'') is shown in red. It has similar performance to the other estimators when
the mass fraction is less than $0.5$ but is worse otherwise, with the performance deteriorating more strongly for higher
mass fractions. 
When the mass fraction is zero, so the failure distribution is absolutely continuous, all methods are identical. 

The right-most pane shows the integrated squared error of the baseline cumulative hazard estimator, where the integral is
across time. The Cox (exact) estimator uses the standard Breslow estimator and the Cox (Efron) estimator uses an Efron-style
modification to the Breslow estimator, although each has different coefficient estimates plugged in. The proposed baseline
cumulative hazard estimator, defined in \Cref{eq:profile-baseline}, has generally the best performance, although with mass fraction
less than around 0.4 all considered estimators are essentially equivalent.

\section{Application: advanced gastric cancer}
\label{sec:data-example}

In this section, we briefly illustrate the methods using the advanced GASTRIC meta-analysis \citep{paoletti2013progression},
available as \texttt{gastadv} in the \texttt{surrosurv} R package \citep{rotolo2018surrosurv}. The data contain 4,069
patients from 20 randomized chemotherapy trials. The endpoint is progression-free survival, defined as the minimum of the
time to detected disease progression and the time to death. 
Progression is generally detected at scheduled clinical assessments, producing discrete failure times, whereas death can
occur and is measured continuously between assessments. 
We treat censoring as conditionally independent. 
Of the 3,820 progression-free survival events, 2,914 were disease progressions and 906 were deaths without a previously recorded progression.

\begin{table}[ht]
\centering
\caption{Estimated chemotherapy effect in the advanced GASTRIC data, adjusted for trial membership, with estimated standard
errors in parentheses.}
\label{tab:gastric}
    \begin{tabular}{lccc}
    \hline
     & Proposal & Cox (exact) & Cox (Efron) \\
    \hline
    Chemotherapy versus control
        & $-0.2240 \,(0.0327)$
        & $-0.2239 \,(0.0337)$
        & $-0.2234 \,(0.0336)$ \\
    \hline
    \end{tabular}
\end{table}

The model covariates we used were a treatment (chemotherapy) indicator and fixed effects for trial number, and we treat the
coefficient of the treatment indicator as the estimand. 
Table~\ref{tab:gastric} compares the proposed estimator, the Cox estimator, and the Efron approximation. 
The three estimators produced nearly identical treatment effect estimates and standard errors. 

Unlike the estimates, the runtime across the methods was quite dissimilar. The Cox estimator was $410\times$ slower than the
Efron estimator, and our manual implementation of the proposed estimator was not comparable. Indeed, in the simulations 
in \Cref{sec:sim}, the Cox estimator could not be computed with $n=1000$ (a
doubled sample size) due to arithmetic overflow despite the manual implementation of the proposed estimator having no
computational issues.

\section{Discussion}
\label{sec:conc}

The Cox regression coefficient $\bbeta$ is among the most widely used estimands in applied work. Its efficient estimation has been
intensely studied when the failure time is absolutely continuous; surprisingly it has been only lightly studied for general
distributions. This work addresses that methodological gap and also introduces a linearized version of the Cox estimator, defined 
using the efficient score, for which the estimating function can be computed faster. Indeed, in the simulation setting of \Cref{sec:sim}, the Cox
estimator \citep[as rigorously implemented by][]{therneau2023package} would generally fail due to numeric overflow for $n=1000$. 
The theory made use of the censoring martingale $M_C^\dagger$ recently studied by \citet{baer2023censoring}
that is generally more suitable in calculations than the Doob-Meyer censoring martingale. 

The Cox model $\mathcal{M}$, defined as a mixture of hazard ratios at continuous failure times and odds ratios at discrete
failure times, is not universally applicable to all research questions. The analysis in this work may open the
door to comprehensive analysis of related estimands in related models. For example, it may be that the distribution at failure
mass points should be variation independent of the distribution off mass points; preliminary work suggests that an
asymptotically efficient estimator of the Cox regression coefficient in this model is the usual Cox estimator after discarding ties. 
Alternatively, there may be a model specification that reduces to the usual Cox model when there are no discontinuities in the
failure distribution, yet remains readily interpretable otherwise, such as in the assumption-lean development from 
\citet{vansteelandt2023assumption}. On the other hand, 
if the observed data is believed to be from a grouped Cox model where the grouping
intervals shrink as the sample size $n \to \infty$, an efficiency calculation with a 
triangular array of probability distributions may reveal that the estimator pretending there are no ties is efficient, 
as in the proposal of \citet{breslow1972hazard};
already \citet{cho2025accurate} established consistency for many estimators and \citet{scheike2007maximum} established that
many estimators have the same asymptotic behavior, under certain grouping asymptotics. 
From the author's perspective, ties should be more than an afterthought in the analysis, especially if theoretical support
for applications with ties are sought.

\section*{Acknowledgements}

The author thanks David Oakes for his encouragement while the project was being formulated. 
ChatGPT assisted in the method's numerical implementation and in proofreading the manuscript.

\bibliographystyle{apalike}
\bibliography{refs}
\end{document}